\documentclass[preprint,showpacs]{revtex4-1}
\usepackage{natbib}
\usepackage{url}
\usepackage{bm}
\usepackage{textcase}
\usepackage{color}
\usepackage{graphicx}
\usepackage{amsmath}
\usepackage{ulem}
\usepackage[colorlinks=true,citecolor=blue]{hyperref} 
\hypersetup{pdftitle={Electronic and magnetic properties of 
                      transition-metal doped ScN for spintronics 
                      applications}}

\hypersetup{pdfauthor={A. Houari}}
\hypersetup{pdfdisplaydoctitle}

\begin{document}
\title{Electronic and magnetic properties of transition-metal doped 
       ScN for spintronics applications}
\author{Fares Benissad}
\affiliation{Theoretical Physics Laboratory, 
             Department of Physics, 
             University of Bejaia, 
             Bejaia, Algeria}
\author{Abdesalem Houari}
\email[corresponding author:]{abdeslam.houari@univ-bejaia.dz}
\affiliation{Theoretical Physics Laboratory, 
             Department of Physics, 
             University of Bejaia, 
             Bejaia, Algeria}
\date{\today} 
\begin{abstract}
Motivated by the ongoing interest in nitrides as materials for 
spintronics applications we have studied effects of doping with magnetic 
transition-metal elements (T=Cr,Mn,Fe,Co and Ni) on the electronic 
properties of semiconducting scandium nitride. Using
  density functional together with the generalized gradient approximation
  (GGA) as well as PBE0r hybrid functional (with different mixing
  of the exact exchange), two different doping amounts
  25\% ($\rm Sc_{0.75}T_{0.25}N$) and 10\% ($\rm Sc_{0.9}T_{0.1}N$)
  have been investigated. This is done in comparison to the reference compound ScN with a strong 
  focus on identifying candidates for half-metallic or semiconducting 
  ferromagnetic ground states. Within GGA, only
  $\rm Sc_{0.75}Cr_{0.25}N$ and $\rm Sc_{0.75}Mn_{0.25}N$ 
  are found to be semiconducting and half-metallic, respectively. The use
  of hybrid functional changes drastically these finding, where $\rm Sc_{0.75}Fe(Co,Ni)_{0.25}N$
  become half-metals and $\rm Sc_{0.75}Cr(Mn)_{0.25}N$ are found both semiconductors.
However, additional calculations assuming antiferromagnetic ordering revealed 
that $\rm Sc_{0.75}Cr_{0.25}N$ is the only compound of this series, 
which prefers an antiferromagnetic (and semiconducting) ground state.
For the lower concentration, $\rm Sc_{0.9}T_{0.1}N$,
similar results have been predicted, and all the doped
nitrides are found to prefer ferromagnetic ground state over an antiferromagnetic one. 

\end{abstract}
%
%
\maketitle 
%
%
\section{Introduction}
\label {I}

The simultaneous exploitation of the charge and spin degrees of 
freedom of electronic materials is at the heart of the rapidly 
emerging field of spintronics. Of particular interest in this field 
are especially two types of materials classes, namely, dilute 
magnetic semiconductors (DMS) and half-metallic ferromagnets (HMF) 
\cite{bowen03,wolf01} with the latter being regarded as good candidates 
for spin injection in magneto-electronic devices \cite{fang02}.

Specifically, Mn-doped III-V semiconducting nitrides and arsenides 
such as (Ga,Al,In)(N,As) have been extensively investigated
\cite{sonoda02,yu02}. However, their use in practical applications 
is hampered by their rather low Curie temperatures. So far, the 
highest Curie temperature of about 173\,K was reported for Mn-doped 
GaN \cite{wang05,jungwirth05}. Yet, it has been argued 
that ferromagnetism above even 300\,K could be possible 
in N-rich growth of Ga(Mn)N \cite{haider05}. Further obstacles to 
applications are the known low solubilities of the magnetic ions 
in the semiconductor matrix due to the different crystal structures  
assumed by the metallic dopants. Finally, spin injection is often 
affected by the resistivity mismatches at semiconductor-ferromagnet 
interfaces \cite{herwadkar05,schmidt00}

In contrast, transition-metal (TM) nitride semiconductors and their 
alloys recommend themselves due to their exceptional optical, 
electro-mechanical, and magnetic properties, which make them 
promising candidates for applications in optoelectronic devices
\cite{houari07,herwadkar05,houari08a,houari10,rajan03}.

Scandium nitride ScN crystallizes in the rocksalt structure with a 
lattice parameter of 4.50 \AA \cite{gall98a,moustakas96}. It has an 
indirect band gap of $\sim 0.9$\,eV 
\cite{albrithen00,albrithen04a,burmistrova13,Conibeer07}.
ScN has been experimentally investigated by a variety of techniques 
\cite{naik13,albrithen02,gall98b,takuechi02}. In addition,
{\it ab initio} calculations have been carried out to investigate 
its electronic structure
\cite{monnier85,neckel76,eibler83,stampfl02,qteish06,lambrecht2000},
confirming the experimental results as the rocksalt crystal structure 
and lattice constant. While some DFT-based calculations using the local 
density approximation found ScN as a semi-metal
\cite{monnier85,neckel76,eibler83,jafar10,tran17}, More elaborated calculations
using the framework of the $GW$ approximation succeeded to reproduce
the experimental band gap of $\sim 0.9 $\,eV
\cite{stampfl02,qteish06,lambrecht2000,rinke05}. 

Considerable effort was devoted to tailoring the physical properties 
of ScN through non-magnetic doping \cite{albrithen04a}. However, only 
few experimental studies have been reported on dilute ferromagnetism 
in doped ScN \cite{albrithen04b,saha13,constantin09,sharma18}. 
Al-Brithen {\it et al.}\ incorporated Mn atoms to obtain
Sc$_{1-x}$Mn$_{x}$N type alloys with $0\leq x \leq 0.25$, crystallizing 
in a tetragonal structure \cite{albrithen04b}. Electronic and optical
properties of Sc(Mn)N thin films were also investigated, with up to
11\% doping \cite{saha13}. The authors suggested that dilute manganese
doping compensates for the high {\it n}-type carrier concentrations,
and an {\it n}-type to {\it p}-type carrier type transition could be
observed.  However, the magnetic properties of these films were not
studied.  Insertion of a low amount of iron (Fe) atoms in ScN has been
achieved by Constantin {\it et al.} \cite{constantin09}, who gave a
magnetic moment of $\sim$ 0.037 $\rm \mu_B$/Fe atom at the
experimental volume of the pristine system.

On the theoretical side, structural and optical properties of $\rm
Cr_{1-x}Sc_{x}N$ alloys have been studied together with Cr-doped GaN
\cite{alsaad10b}. However, with respect to the magnetic properties,
focus was on the latter material. Supercell calculations on Mn-doped
ScN have been carried out to study trends of the magnetic properties
\cite{herwadkar05,herwadkar08,alsaad10a,houari08b}. It was found that
the exchange interactions are long range and affected by disorder and
carrier concentrations \cite {herwadkar08,alsaad10a}.
Furthermore, a structural phase transition has been predicted in Mn$_x$Sc$_{1-x}$N
from a zinc-blende phase to a hexagonal phase \cite{alsaad10a}.  For iron
doping, Fe$_{0.25}$Sc$_{0.75}$N has been investigated by means of DFT
calculations, which predicted a magnetic semiconductor \cite{sharma18}. 
The study, however, has focused more on the transport and thermoelectric
properties.

Recently, a DFT-based study using the GGA$ +U $ method has been reported 
on a large series of transition-metal doped (T,Sc)N (with T=Ti, V, Cr, 
Mn, Fe, Co and Ni)\cite{sukkabot2019}. The occurrence of half-metallicity 
was predicted for V-, Cr-, and Mn-doping, whereas metallic behavior was
expected for doping with late transition-metal atoms.

Extending a previous study on pure ScN and $\rm Sc_{0.75}Mn_{0.25}N$ 
as based on the GGA \cite{houari08b}, we investigate in the present 
investigation ScN doped with a larger variety of transition metal 
including Cr, Mn, Fe, Co, and Ni). In doing so, we go beyond previous 
studies by applying a hybrid-functional approach in addition to GGA
calculations.

\section{Theory and Computational Method}
\label{paw}

The first principles calculations of the present work are based on
density functional theory (DFT) \cite{hohenberg64,kohn65}. In a 
first step, exchange-correlation effects are accounted for by the
generalized gradient approximation (GGA) \cite{perdew96a}. In 
addition, we applied a hybrid functional approach, which as well 
known replaces part of the exchange functional of the GGA by the 
exact exchange term provided by the Hartree-Fock method
\cite{becke93,perdew96b}. Within this general approach, several 
implementations are in use. While the PBE0 method includes the 
exact exchange with a fraction of 25\% in all space \cite{adamo99}, 
the HSE06 method uses this fraction only for the short-range part 
while keeping the long-range part of the exchange interaction at 
the GGA level \cite{heyd03}. This idea is put to its extreme by 
the PBE0r functional, which restricts the exact exchange interaction 
to onsite matrix elements in a local orbital basis 
\cite{bloechl11,bloechl13}. The accuracy of the PBE0r
  functional, with respect to experiment, has been demonstrated in a
  study of transition-metal oxides \cite{sotoudeh17}. Furthermore,
  a very recent extensive benchmark investigation using HSE06 as well as
  PBE0r functionals has been reported \cite{eckhoff20}, where the
  the latter is found to have an accuracy in the same range of the former.
The PBE0r method is also used in the present study, where
the exact exchange contribution is considered with different fractions
(see below in \ref{sctn}).
Notice that a small value of 5\% suffices to reproduce the experimentally
reported ScN band gap of 0.9~eV.

The electronic and magnetic properties were calculated using the 
CP-PAW code, which is the original implementation of the Projector 
Augmented Wave (PAW) method \cite{bloechl94a}. It uses the 
Car-Parrinello {\it ab initio} molecular dynamics (AIMD) framework 
to simultaneously calculate the atomic structure and the electronic 
wave functions \cite{car85}.The Mermin functional
  \cite{mermin65} was applied to treat variable occupations of the
  one-electron energy eigenstates, allowing to describe correctly
  metallic systems.

Total energy convergence was assumed when the difference between two
successive iterations was less than 10$^{-5}$\,Hartree. The plane-wave
cutoff was 40\,Hartree and the Brillouin-zone integrations were performed
on a ($6\times6\times6$) mesh (leading to 260 k-points in
the Brillouin-zone)
using the linear tetrahedron method
\cite{jepsen71,lehmann72} with Bl\"ochl correction
\cite{bloechl94b}. The convergence with respect to
  plane waves expansion as well as k-points, has been carried out
  carefully by increasing the cutoff value until an accuracy of 
  (i.e the variation of the total energy is less than)
   10$^{-5}$\,Hartree is reached. Full description of the
  calculations details can be found in our recent publication
  \cite{houari19}. In addition, all structural parameters have been optimized. The
  relaxed atomic positions (geometry optimization) and the electronic ground state are
  obtained efficiently using a friction, within the Car-Parrinello molecular
  dynamics method. The lattice constant corresponding to theoretical
  equilibrium, is optimized by an isotropic expansions and
  contractions.

\section{Results and Discussions}
\label{results}

\subsection{Electronic structure of pure ScN}
\label{scn}

Since the electronic properties  
of $\rm ScN$ are well established both experimentally and theoretically, 
this section just serves as a reference for the subsequent discussion.
More thorough description can be found in literature, as mentioned above
in the introduction. The equilibrium structure of pure ScN has been determined by minimization 
of the total energy versus unit cell volume using both the GGA and hybrid
functionals. From this, lattice constants of 4.58\,\AA\ and 4.52\,\AA\ were 
obtained from the GGA and the hybrid-functional calculations, respectively. 
While the former as expected is somewhat larger than the experimental 
value of 4.50\,\AA, the latter shows better agreement. 

\begin{figure}[ht]
\begin{center}
\includegraphics[width=0.8\linewidth]{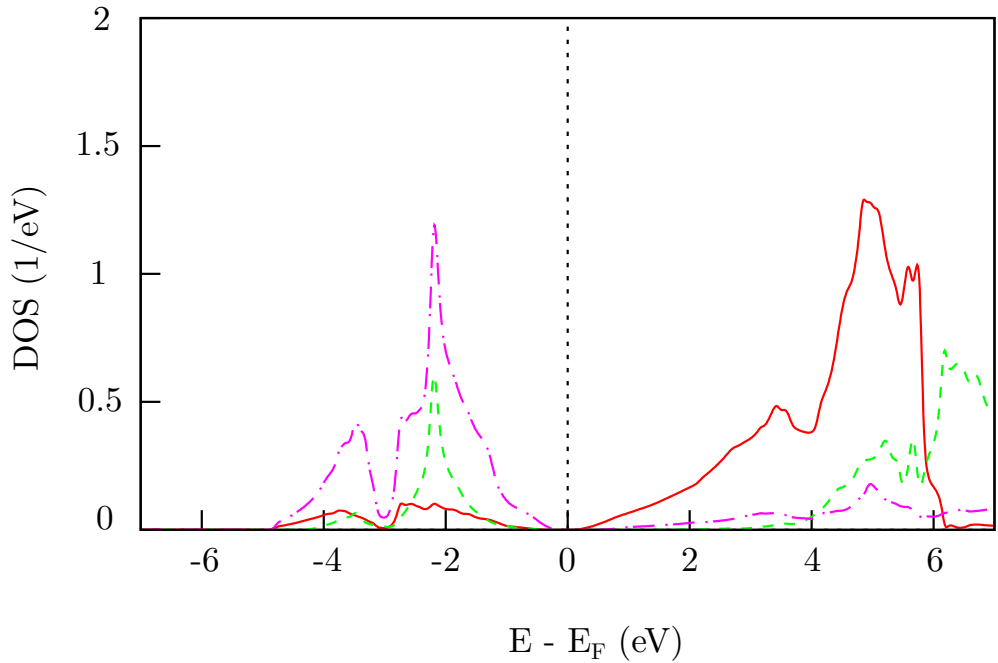}
\caption{(Color online) Site- and orbital projected densities of states 
         (DOS) of $\rm ScN$. Sc-$ d $ $ t_{2g} $, Sc-$ d $ $ e_g $, and 
         N $ p $ contributions as given in red (solid), green (dashed), 
         and magenta (dash-dotted), respectively.}  
\label{scn_dos}
\end{center}
\end{figure}

The site- and orbital-projected densities of states as calculated using 
the GGA, within the experimental lattice, is displayed in Fig.~\ref{scn_dos}.
they confirm the semiconducting state with a band gap of about
0.35\,eV, which is much smaller than the measured experimental one
($\sim$0.9 eV). This is due to the well-known failure of the local and
semi-local approximations to reproduce experimental band
gaps. Since many previous LDA and GGA calculations
  have predicted vanishing gap, as pointed out in the
  introduction\cite{jafar10,tran17}, we have performed additional
  complementary GGA with a another method based on a different wavefunction
  basis set. Using the full-potential augmented spherical wave method
  (FP-ASW) \cite{eyert00,eyert13}, with a carefully converged
  calculation, a band gap of ~0.38 eV has been computed. Still being
  far from the experimental one, this result is close to our first
  one.

While the nitrogen 
{\it 2p}-states are more prevailing in the valence band, the conduction 
band is dominated by the transition-metal {\it d} states. The latter 
fall into non-bonding {\it t$_{2g}$} sub-bands in the lower part 
(with a large peak around 5\,eV), whereas the {\it e$_g$} states due 
to stronger overlap with the N-{\it 2p} and, hence, larger splitting 
into bonding and anti-bonding manifolds are found at higher energies. 
As already reported, the trivalent character of Sc is almost complete,  
i.e.\ ScN can be written as Sc$^{+\delta}$N$^{-\delta}$ with 
$\delta \simeq 3$, making the compound almost ionic \cite{houari08b}. 

In a second step, hybrid-functional calculations were performed with 
a 5\%-admixture of the exact exchange interaction as mentioned above, 
which allowed to reproduce the experimentally observed band gap of 
0.9\,eV.

\subsection{Trends of electronic structure and magnetic properties of
  Sc$_{1-x}$T$_x$N (T=Cr, Mn, Fe, Co, Ni)}
\label{sctn}

This section is devoted to the study the effect of the substitution
  of scandium by several transition metal atoms, i.e. $\rm Sc_{1-x}T_xN$, where
T=Cr, Mn, Fe, Co, and Ni, with two different x=0.25 and x=0.10 concentrations.

In the first case, the simulations are performed 
using a simple-cubic supercell of the fcc-rocksalt structure of the 
form Sc$_3$T$_1$N$_4$. The calculated theoretical equilibrium lattice
parameters are found very close to each others in the whole series.
With hybrid functional framework, we computed the following values:
4.423\,\AA\ (T=Cr), 4.439\,\AA\ (T=Mn), 4.457\,\AA\ (T=Fe),
4.430\,\AA\ (T=Co) and 4.445\,\AA\ (T=Ni), whereas those calculated
by GGA are about 3\% larger. Thus, the substitution leads a slightly
smaller lattice constants compared to pure ScN. We mention, however,
that neither experimental nor theoretical data are available in
literature, to which our results could be compared. 

\begin{figure}[h]
\begin{center}
\includegraphics[width=0.4\linewidth,height=20cm]{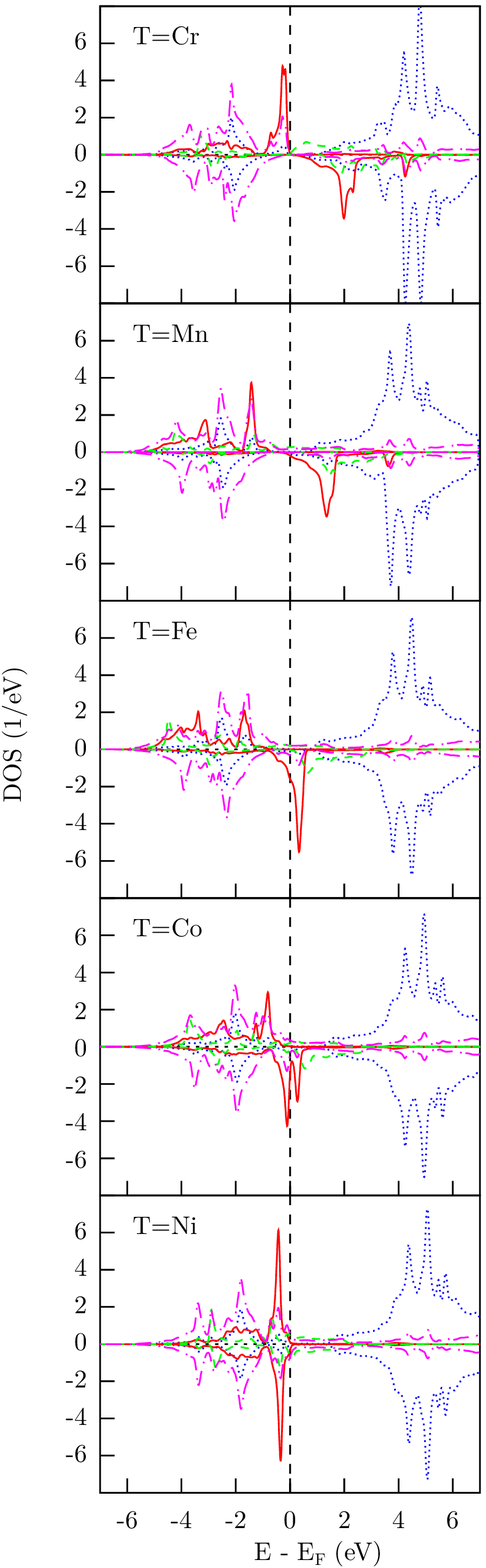}
\caption{(Color online) Site-, orbital-, and spin-projected density of 
          states (DOS) of $\rm Sc_{0.75}T_{0.25}N$ with T=Cr, Mn, Fe, 
          Co, and Ni as obtained within the GGA. 
          Sc-{\it d}, T-{\it d}-t$_{2g}$, T-{\it d}-e$_g$, and N-{\it p} 
          contributions are given in blue (dotted), red (solid), green 
          (dashed), and magenta (dash-dotted), respectively.}
\label{sc3t1n4_ggados}
\end{center}
\end{figure}

We have first considered a ferromagnetic alignment of
the transition metal atoms. The ferromagnetic 
state is found more stable than the spin-degenerate situation for all 
five dopants considered here. The electronic densities of states of 
the five compounds as described within the GGA framework are shown 
in Fig.~\ref{sc3t1n4_ggados}. 
Obviously, all types of substitution lead to loss of the semiconducting 
character of pure ScN. While Mn, Fe, Co, and Ni insertion induce metallic 
behavior, $\rm Sc_{0.75}Cr_{0.25}N$ is found to be a half-metal. In the 
latter case, an energy band gap of $\sim$ 0.9\,eV is opened in the 
spin-down states, whereas a small contribution to the  density of states 
at the Fermi level can be seen in the spin-up states. Hence, the compound 
is at the verge of being a semiconductor. Similarly, 
$\rm Sc_{0.75}Mn_{0.25}N$ is close to being half-metallic. 

In all the compounds, the distribution of the scandium {\it d}-states 
remains almost unchanged to that in pure ScN indicating only weak 
hybridization with the {\it d} states of the dopant. In contrast, 
the {\it d} states of the substitutional alloying elements can be 
regarded as in-gap states between N-{\it p} and Sc-{\it d} states of ScN.

In $\rm Sc_{0.75}Cr_{0.25}N$ and $\rm Sc_{0.75}Mn_{0.25}N$ the 
non-bonding {\it d}-t$_{2g}$ states of chromium and manganese 
experience a strong exchange splitting. As a result, spin-majority 
states are completely filled leading to a high-spin configuration 
of the Cr and Mn t$_{2g}$ shells. However, the two compounds differ 
with respect to the {\it d}-e$_g$ states. Like in ScN, the dopant 
{\it d}-e$_g$ states hybridize strongly with the N-{\it p} states 
and are split into bonding and anti-bonding manifolds below and 
above the Fermi energy, respectively. While in the case of Cr the 
latter are empty for both spin populations, the Mn-{\it d}-e$_g$ 
states are only half-filled. As a result, magnetic moments per Cr
and Mn atom of 2.9$\rm \mu_B$ and 3.8$\rm \mu_B$, respectively,
were obtained in the doped compounds. 

Beyond Cr and Mn, the exchange splitting of the t$_{2g}$ states 
decreases progressively due to the filling of the corresponding 
minority-spin states, which are shifted below the Fermi level. 
Finally, for $\rm Sc_{0.75}Ni_{0.25}N$ the exchange splitting is 
almost vanishing and both the majority and minority spin channels 
of the Ni-t$_{2g}$ states are completely filled. The calculated
magnetic moments per dopant atom are 2.8$\rm \mu_B$, 1.7$\rm \mu_B$, 
and 0.2$\rm \mu_B$ for the Fe-, Co-, and Ni-doped compounds, 
respectively.
 
\begin{figure}[h]
\begin{center}
\includegraphics[width=\linewidth,height=20cm]{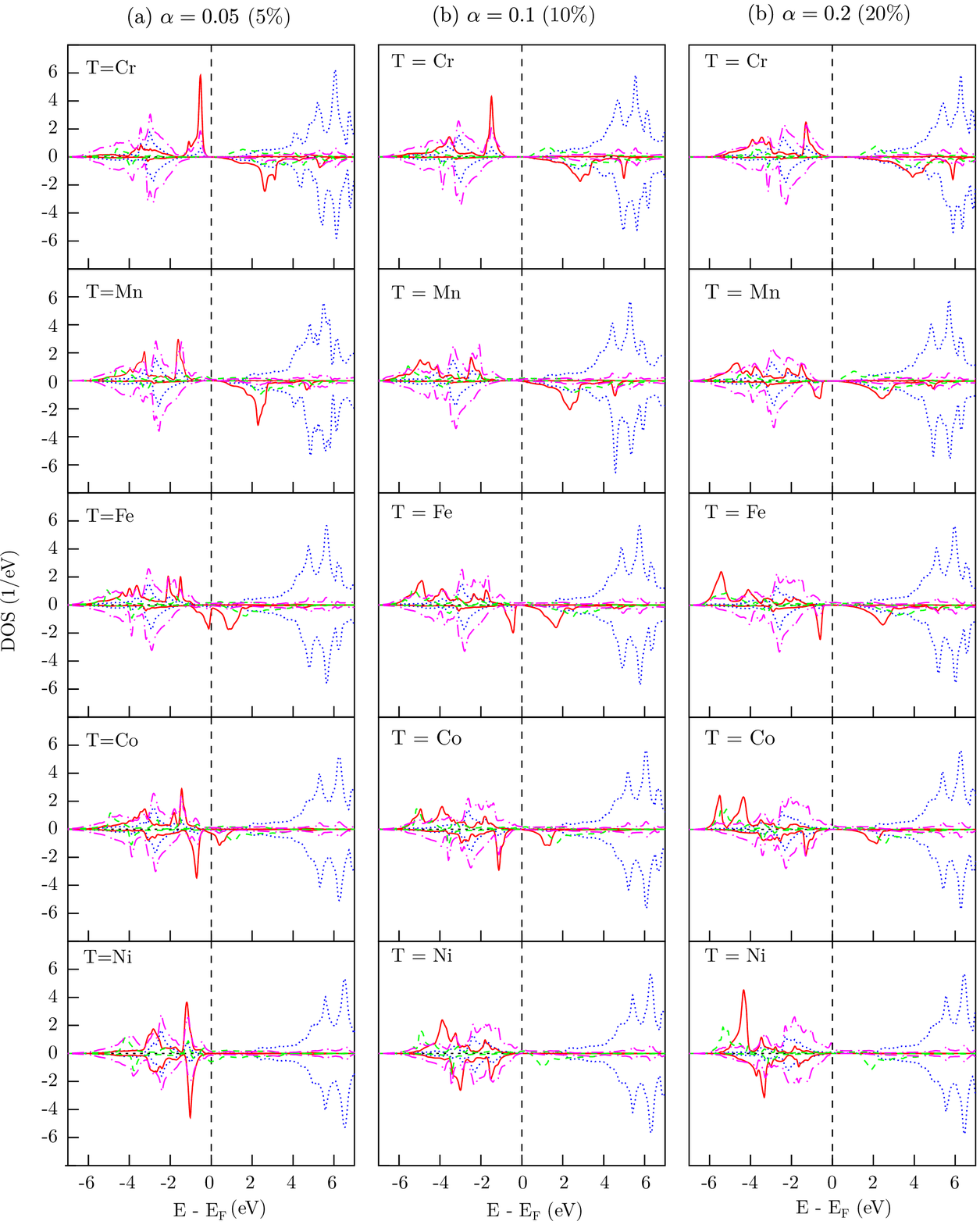}
\caption{(Color online) Site-, orbital-, and spin-projected density of 
          states (DOS) of $\rm Sc_{0.75}T_{0.25}N$ with T=Cr, Mn, Fe, 
          Co, and Ni from hybrid-functional calculations 
          with three mixing parameters (admixture in \%) of the
          exact exchange: $\alpha=$0.05
          (5\%), $\alpha=$0.1 (10\%) and $\alpha=$0.2 (20\%). The 
          Sc-{\it d}, T-{\it d}-t$_{2g}$, T-{\it d}-e$_g$, and N-{\it p} 
          contributions (color and lines) are the same as in
          Fig. \ref{sc3t1n4_ggados}.}
\label{sc3t1n4_hybdos}
\end{center}
\end{figure}

Since the main purpose of this investigation is the search for
ferromagnetic half-metals and semiconductors, a correct evaluation of
the band gap is for great importance. As mentioned above, use of the
hybrid functional allowed to obtain the experimentally reported band
gap for ScN. In order to explore the effect of the
  main ingredient of the hybrid functional scheme which is the
  inclusion of the exact exchange, we have performed calculations with
  three different mixing parameters (i.e the weight of the Fock term,
  noted $\alpha$). The PBE0r functional allows to specify this
  mixing parameter on each site. So, we kept the one on Sc and N atoms
  fixed to 5\% (or $\alpha=0.05$, like in pure ScN), while increasing the
  one on the transition metal element gradually: 5\% ($\alpha=0.05$), 10\%
  ($\alpha=0.10$) and finally 20\% ($\alpha=0.20$).  The corresponding
  results of the site- orbital-, and spin-projected densities of
  states for the five doped compounds are displayed in
  Fig.~\ref{sc3t1n4_hybdos}. 

  In Fig.~\ref{sc3t1n4_hybdos}-(a) corresponding to
  the lowest value of the mixing parameter ($\alpha=0.05$),
  the electronic structure of $\rm Sc_{0.75}(Fe,Co,Ni)_{0.25}N$ doped nitrides 
  remains almost unchanged, with respect to GGA, and still shows metallic
  character. The situation is different for the doping with Cr and Mn.
  Specifically, for $\rm Sc_{0.75}Cr_{0.25}N$ a ferromagnetic
  semiconducting is found with a band gap of $\sim$0.5 eV, whereas
  $\rm Sc_{0.75}Mn_{0.25}N$ turns out to be a ferromagnetic half-metal
  with a finite DOS in the majority-spin channel and a band gap of
  $\sim$0.8 eV in the minority-spin channel.

 When increasing the the mixing parameter to $\alpha=0.1$,
  the picture start to change noticeably, as shown in Fig.~\ref{sc3t1n4_hybdos}-(b).
  Here, while the electronic character of two first $\rm Sc_{0.75}Cr(Mn)_{0.25}N$
  compounds remains the same as for $\alpha=0.05$ (just with a slightly enhanced
  gap fo the Cr-doped nitride $\sim$0.7 eV), we notice that $\rm Sc_{0.75}Fe(Co)_{0.25}N$
  turn to be half-metallic ferromagnets with a small minority spin
  band gap. Moreover, Ni-doped nitride
  $\rm Sc_{0.75}Ni_{0.25}N$ is found at the verge of being a half-metal too,
  where a tiny gap starts to open in the minority spin channel. This situation
  becomes completely resolved with the highest mixing parameter value $\alpha=0.2$,
  leading to a clear semiconducting ferromagnetism in the two first compounds
  $\rm Sc_{0.75}Cr(Mn)_{0.25}N$, (with band gap values of $\sim$0.85 and
  $\sim$0.45, for Cr and Mn doping, respectively). In the other Fe-, Co- and
  Ni-doped nitrides, a very net half-metallic character is stabilized.
  Of course, both physical states are of high interest for spintronics 
 applications. 

  Overall, the trend of the densities of states of all the compounds is
  very similar to the one observed for the GGA calculations, for the
  three mixing parameters, especially concerning the Sc and N electronic DOS. In
particular, the projected DOS given in Figs.~\ref{sc3t1n4_ggados} and
\ref{sc3t1n4_hybdos} have almost the same shape and distribution and
the general discussion of the GGA results applies also to the
hybrid-functional calculations apart from the relative shifts of the
groups of bands. There is, however, a noticeable difference
  concerning the exchange splitting of t$_{2g}$ and e$_g$ shells in
  the transition metal ions, when increasing the weight of the
  Fock term $\alpha$. This can be seen from the computed values
  of the magnetic moments for the different cases, as shown in
  Tab. \ref{tab:table1}. A saturated magnetic moments for all the compounds
  are obtained with the two last fractions of the exact exchange (i.e
  $\alpha=0.1$ to $\alpha=0.2$), where semiconducting or half-metallic
  behavior are obtained. For the lowest value $\alpha=0.05$, however,
  the metallic character of Fe-, Co- and Ni-doped nitrides leads to
 non-integer magnetic moments. Hence, as expected, the hybrid-functional
approach favors a high-spin configuration as compared to the GGA.
We notice that within both frameworks the total magnetic moment per unit
cell arises almost exclusively from the dopant atom. This is the case
in all five compounds, where the $N$ atom is found to carry a moment
of less than 0.1\,$\mu_B$.

  According to the densities of states of Fig.
  \ref{sc3t1n4_hybdos}, the magnetic moments of all 
doped compounds arise almost completely from the t$_{2g}$ states
(visible from the exchange splitting between the two spin populations),
except the Co- and Ni-doped ones. In the latter, in particular, a
small contribution from e$_g$ states is visible.  
This is a consequence of the weak hybridization of these states with 
the nitrogen $p$-states, whereas the e$_g$ states due to their strong 
bonds with the nitrogen states and the resulting large bonding-antibonding 
splitting showing very similar spin-up and spin-down contributions.

\begin{table*}[ht]
  \caption{\label{tab:table1} Calculated magnetic moments, in $\mu_B$, of the
  transition metal elements within GGA as well as hybrid functionals.}
\begin{ruledtabular}
\begin{tabular}{c|c|ccc} 
$ {\rm {Sc_{0.75}T_{0.25}N}} $ & GGA &                           & Hyb. Func    &   \\
\hline
                              &    & $\alpha=0.05$              & $\alpha=0.1$ &  $\alpha=0.2$ \\
\hline
$ {\rm T = Cr} $   & 2.9  & 3    & 3 &  3\\
$ {\rm T = Mn} $   & 3.8  & 4    & 4  & 2\\
$ {\rm T = Fe} $   & 2.8  & 3.5  & 3 &  3\\
$ {\rm T = Co} $   & 1.7  & 1.9  & 2 & 2 \\
$ {\rm T = Ni} $   & 0.2  & 0.85 & 1 & 1\\

\end{tabular}
\end{ruledtabular}
\end{table*}

The particular case of $\rm Sc_{0.75}Fe_{0.25}N$ has been studied by
Sharma {\it et al.}\ using the modified Becke-Johnson local-density
approximation (mBJ-LDA) \cite{tran09}. They have found that the 
{\it p-d} hybridization between iron and nitrogen states gives rise 
to a magnetic semiconductor, which is in contradiction with our
results, that is a metallic character predicted from GGA and half-metallic
one with hybrid-functional calculations.

Very recently, Sukkabot {\it et al.}\ using the GGA$+U$ method with a 
$U$ value of 3.9 eV for all the compounds have investigated ScN doped 
with several transition elements \cite{sukkabot2019}. They found 
half-metallic ferromagnetism in both Mn- and Cr-doped ScN.
  While the former result is in agreement with our study
  with a low admixture of exact exchange ($\alpha=0.05$), we predict
  in contrast that Cr-doped ScN is a semiconductor, as well as
  Mn-doped one with larger mixing parameter.

To finalize the investigation of $\rm Sc_{0.75}T_{0.25}N$ compounds,
the calculations of the ferromagnetic states were complemented 
by investigations regarding possible antiferromagnetic alignment of 
the magnetic moments of the dopant atoms. To this end, supercells
(formed by doubled simple-cubic cell) were considered to simulate  
a simple type I antiferromagnetic alignment (AFM$\langle001\rangle$).
The latter consists on a ferromagnetic plans alternating along the z-axis
direction. To the best of our knowledge, the study of AFM ordering
in doped-ScN has not been reported elsewhere. 

\begin{figure}[h]
\begin{center}
\includegraphics[width=0.5\linewidth,height=10cm]{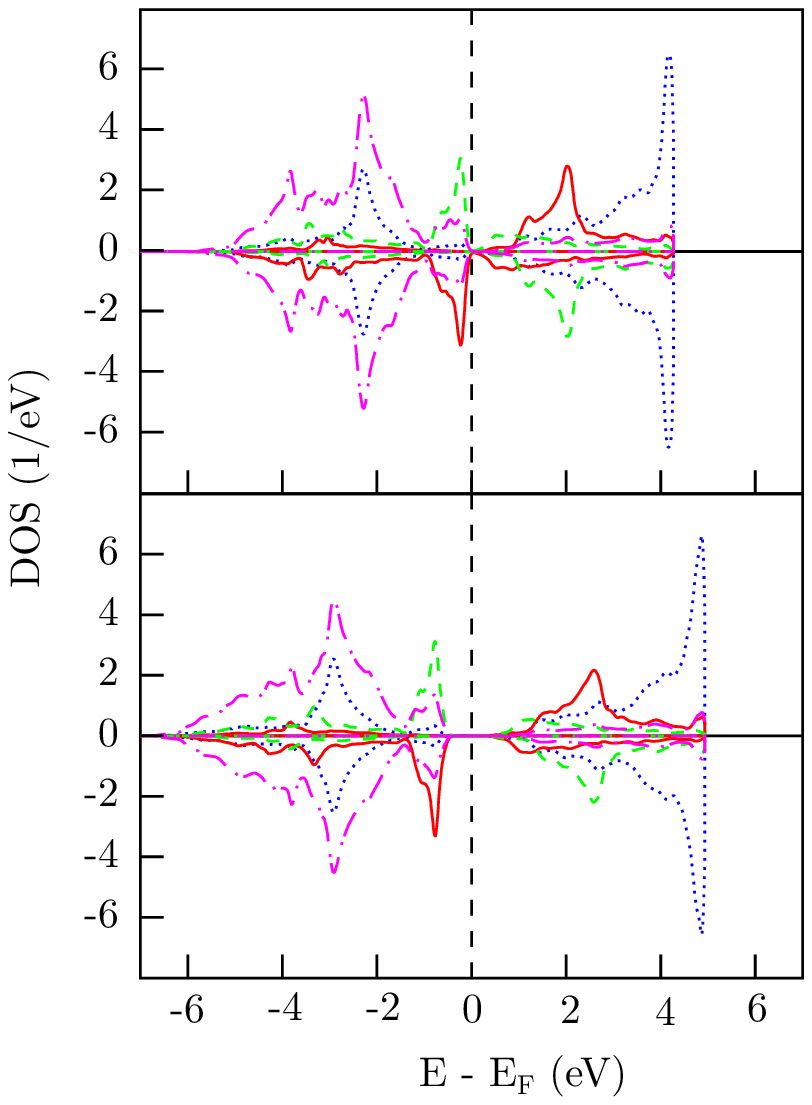}
\caption{(Color online) Site-, orbital-, and spin-projected density 
          of states (DOS) of antiferromagnetic $\rm Sc_{0.75}Cr_{0.25}N$ 
          as obtained from GGA calculations (top) and hybrid-functional 
          calculations (bottom). 
          Sc-{\it d} and N-{\it p} contributions are given in blue (dotted) 
          and magenta (dash-dotted), respectively, while the contributions 
          of the Cr-{\it d} states of the two sublattices are given in red 
          (solid) and (green) (dashed).} 
\label{sccr_afm_dos}
\end{center}
\end{figure}

As a result, except for chromium doped ScN antiferromagnetic order 
was found to be less stable than the ferromagnetic state discussed 
above. Specifically, in $\rm Sc_{0.75}Cr_{0.25}N$ the Cr atoms prefer 
to antiferromagnetic over ferromagnetic one, which is by $\sim 8$\,mHa
and $\sim 10$\,mHa per formula unit higher in energy, in GGA and hybrid
functional, respectively. 
The site-, orbital, and spin-projected densities of states of 
antiferromagnetic $\rm Sc_{0.75}Cr_{0.25}N$ are shown in 
Fig.~\ref{sccr_afm_dos}. 
The GGA calculations lead to a metallic state with an almost vanishing 
density at Fermi level and magnetic moments of $\pm$2.75\,$\rm \mu_B$ 
located at the Cr ions. In contrast, use of the hybrid functional gives 
rise to a semiconducting state with a band gap of $\sim 0.8$\,eV, and a 
slightly enhanced Cr magnetic moment of $\pm $2.8\,$\rm \mu_B$.

\begin{figure}[h]
\begin{center}
\includegraphics[width=\linewidth,height=20cm]{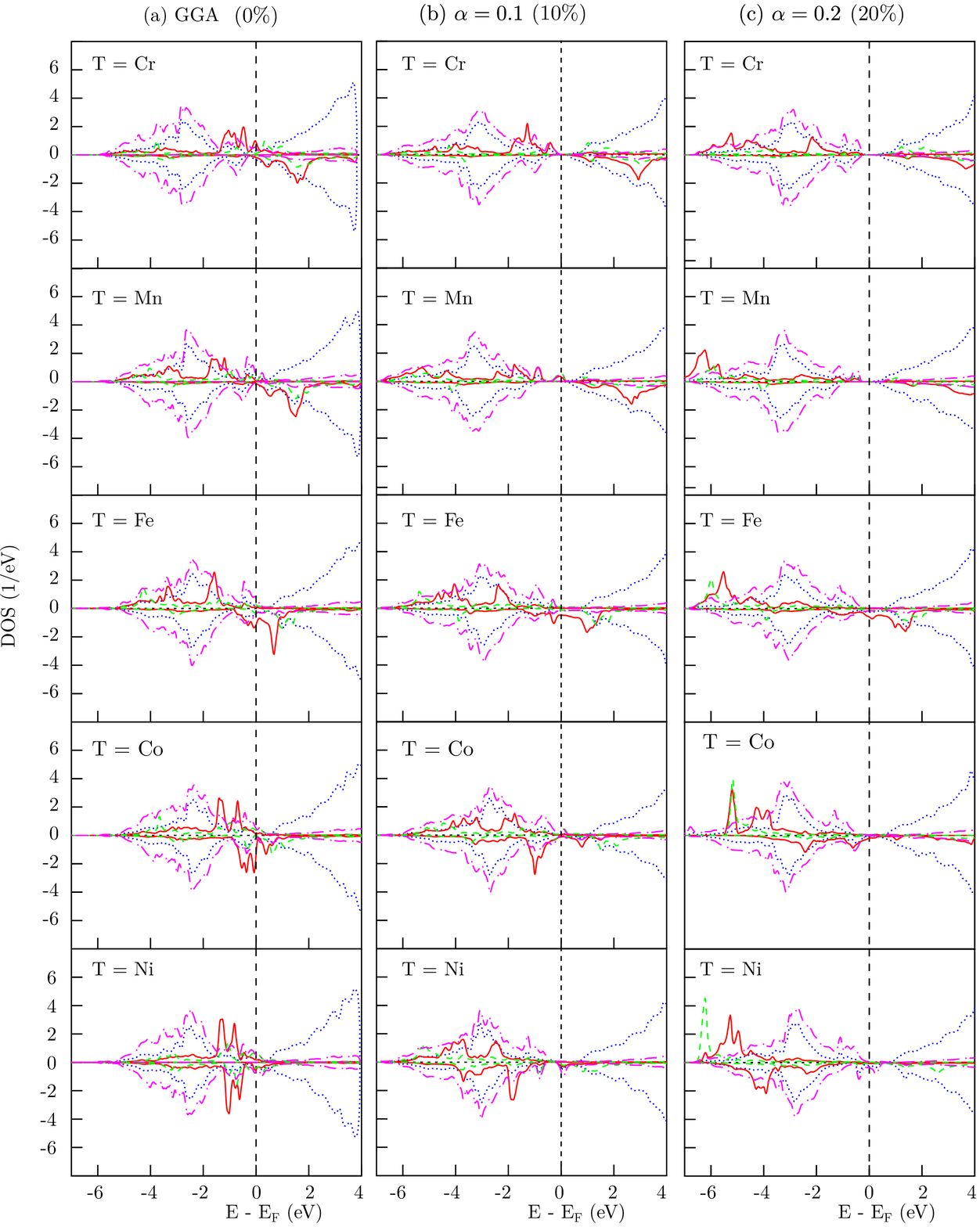}
\caption{(Color online) Site-, orbital-, and spin-projected density of 
          states (DOS) of $\rm Sc_{0.9}T_{0.1}N$ with T=Cr, Mn, Fe, 
          Co, and Ni obtained from GGA as well as hybrid-functional
          with $\alpha=$0.1 and $\alpha=$0.2.
          The contributions of the atomic
          species (lines and colors) are the same as in Fig. \ref{sc3t1n4_hybdos}.}
\label{t1sc9_alldos}
\end{center}
\end{figure}

 The second part of our study consists to investigate
  the electronic structure and magnetic properties of the doped-ScN nitride
  with a lower substitution concentration, that is $\rm Sc_{0.9}T_{0.1}N$
  ordered compounds (T=Cr, Mn, Fe, Co and Ni). Like in the previous case,
  we started first by considering a ferromagnetic alignment of the transition
  metal ions. The obtained densities of states within GGA are shown along side
  with those obtained with hybrid functional frameworks in Fig.\ref{t1sc9_alldos}.
  Here, we show the results of moderate ($\alpha=0.1$) and
  highest ($\alpha=0.2$) mixing parameters, and we omit the lowest
  one ($\alpha=0.05$) since it does not bring noticeable changes to
  the GGA findings. 

  While GGA describes all the compounds
  as metals, the hybrid functional predicts both semiconducting and half-metallic
  ferromagnetic characters. Let us first notice that the densities of
  N and Sc atoms are nearly not affected with respect to pure ScN, where the
  states of the former dominate the valence band and those of the latter
  are prevailing in the conduction band. In GGA (Fig. \ref{t1sc9_alldos}-a),
  the transition metal $\rm t_{2g}$ states experience an exchange splitting
  of the two spin populations due to the weak bonding with nitrogen, which decreases in intensity
  going from Cr to Ni. In contrast, the $\rm e_g$ states with a strong hybridization
  with N states show similar spin-up and spin-down distributions. The hybrid functional
  with a mixing parameter of $\alpha=0.1$ leads to a half-metallicity in Cr-, Mn-,
  Fe- and Ni-doped nitrides; whereas the Co-doped one still behaves as a metal. When
  increasing the parameter to $\alpha=0.2$, some changes occur and we find a
  similar situation to the high concentration (x=0.25) compounds.
  Thus, Cr- and Mn-doping lead to semiconducting nitrides, while
  $\rm Sc_{0.9}Fe(Co,Ni)_{0.1}N$ ones become half-metals. We notice that the
  overall trends of the transition metal densities of states
  and the contribution to the magnetic moments are also similar to
  the high concentration case.

  Finally, we complemented this second part of the study
  by a check on the possible stability of antiferromagnetic alignment
  of the transition metal ions (AFM$\langle001\rangle$, as in the first case of high
  concentration). Here, and in contrast to the former case, all the $\rm Sc_{0.9}T_{0.1}N$
  doped nitrides are found energetically more stable in the ferromagnetic configuration.
%
%

\section{Summary and Conclusion}
\label{conclusion}

To conclude, first-principles calculations as based on density functional 
theory were used to explore the change in electronic and magnetic behavior 
of ScN upon substitution with transition-metal ions T=Cr, Mn, Fe, Co, and Ni,
i.e. $\rm Sc_{1-x}Cr_xN$, with two concentrations x=0.25 and x=0.1. 
In particular, the influence of exchange effects beyond the semi-local 
approximation as captured by the generalized gradient approximation was 
investigated. 

In a first step, the electronic structure of ScN was studied by both GGA 
and hybrid-functional calculations to establish a reference for the doped 
compounds. With a fraction as small as 5\% of the exact exchange functional 
the experimental band gap of ScN of ~0.9\,eV could be reproduced. 

  Within GGA approximation, and initially focusing on
  the ferromagnetic state, we found $\rm Sc_{0.75}T_{0.25}N$ to be
  ferromagnetic metals for T = Mn, Fe, Co, and Ni; whereas $\rm
  Sc_{0.75}Cr_{0.25}N$ is a half-metal. Decreasing the substitution
  concentration to 10\%, all $\rm Sc_{0.9}T_{0.1}N$ compounds are
  found metallic.  Using hybrid-functional framework, with increasing
  exact exchange mixing parameter, the electronic behavior of the
  compounds changes progressively. As a consequence, $\rm
  Sc_{0.75}T_{0.25}N$ are established as ferromagnetic half-metals for
  T = Fe, Co, and Ni, while both $\rm Sc_{0.75}Cr(Mn)_{0.25}N$ are
  turned out to be semiconducting ferromagnets. Similar findings have
  been obtained in second substitution concentration $\rm
  Sc_{0.9}T_{0.1}N$.
  
  Considering antiferromagnetic ordering, it was found unstable for
  all $\rm Sc_{0.75}T_{0.25}N$ compounds except $\rm Sc_{0.75}Cr_{0.25}N$,
  leading to a semiconducting antiferromagnetic ground state of this compound.
  In the lower concentration, however, all $\rm Sc_{0.9}T_{0.1}N$ are found
  to prefer a ferromagnetic ground state.

  We conclude that in the two limits of substitution concentrations,
  the predicted physical properties could be interesting for practical applications
  in spintronics.

\bigskip

\begin{acknowledgements}
The authors gratefully acknowledge V.~Eyert for
  a critical reading of the manuscript and fruitful 
discussions. A.~H.\ also acknowledges P.~E.~Bl\"ochl for providing 
his CP-PAW code.

\end{acknowledgements}

\end{document}